\documentclass[pra,twocolumn,longbibliography]{revtex4-2}
\usepackage{amsmath}
\usepackage{amssymb}
\usepackage{graphicx} 
\usepackage[usenames, dvipsnames]{color}
\usepackage[colorlinks=true,citecolor=blue,urlcolor=purple]{hyperref}

\newcommand{\ket}[1]{\vert #1 \rangle}
\newcommand{\bra}[1]{\langle #1 \vert}

\newcommand{\ketbra}[2]{\vert #1 \rangle \langle #2 \vert}
\newcommand{\abs}[1]{\vert #1 \vert}

\newcommand{\mean}[1]{\langle #1 \rangle}

\renewcommand{\vec}[1]{\pmb{\rm #1}}
\usepackage{xcolor}
\newcommand{\red}[1]{{\color{red} #1}}
\newcommand{\new}[1]{{\color{red} #1}}

\renewcommand{\new}[1]{{ #1}}
\renewcommand{\red}[1]{{ #1}}
\begin{document}

\title{The emergence of classical stochasticity}

\author{Xuan Du Trinh}
\email{xtrinh@cs.stonybrook.edu}
\affiliation{Stony Brook University, Stony Brook, 11794 New York, USA}

\author{Isma\"el Septembre}
\email{ismael.septembre@uni-siegen.de}
\affiliation{Naturwissenschaftlich--Technische Fakult\"{a}t,
Universit\"{a}t Siegen, \\ Walter-Flex-Stra{\ss}e 3, 57068 Siegen, Germany}

\author{Hai-Chau Nguyen}
\email{chau.nguyen@uni-siegen.de}
\affiliation{Naturwissenschaftlich--Technische Fakult\"{a}t,
Universit\"{a}t Siegen, \\ Walter-Flex-Stra{\ss}e 3, 57068 Siegen, Germany}

\begin{abstract}
    We examine the emergence of classical stochasticity for a quantum system governed by a Pauli-type master equation. It is well-known that while such equations describe the evolution of probabilities, they do not automatically justify reasoning based on the assumption that the system exists in a definite state at intermediate times. On the other hand, we show that this assumption is crucial for the standard calculation of stochastic times such as the persistent time and the time of first arrivals.
    \new{It is also curious to point out that the standard derivation of the Pauli master equation based on the Fermi Golden Rule actually proceeds by assuming that the system remains in a definite state as the starting point to derive the transition rate to a different state.
    By studying the so-called ultradecoherence limit, we show that the Pauli master equation can also be arrived at without such an assumption. Instead, that the system exists in a definite state has to be justified by a different argument.}
    We then consider examples of single particles, bosons, and fermions in the ultradecoherence limit to illustrate how classical stochasticity may emerge from quantum mechanics.
\end{abstract}

\maketitle

\section{Introduction}

During the early development of quantum mechanics, Pauli proposed that under certain conditions, quantum dynamics can be (approximately) described by a stochastic process~\cite{Pauli1928}, where the system occupies a specific quantum state $\ket{\mu}$ and jumps to another $\ket{\nu}$ with a certain rate $W_{\nu \mu}$. 
Based on results from Dirac's theory of Fermi's Golden Rule~\cite{Dirac1927EmissionAbsorption}, Pauli gives a concrete example of the computation of the transition rates $W_{\nu \mu}$.
He then derived the evolution equation of the probability distribution \new{$P_\mu$} of the system over the states as
\begin{equation}
    \frac{d P_{\mu}}{d t} = \sum_{\nu \ne \mu} (W_{\mu \nu} P_{\nu} - W_{\nu \mu} P_{\mu}),
    \label{eq:Pauli-master}
\end{equation}
now known as Pauli's master equation~\cite{Pauli1928}. 
In essence, Pauli anticipated an emergent classical picture augmented by stochasticity from the quantum theory, newly founded by Heisenberg and Schr\"odinger~\cite{heisenberg_ber_1925,born_zur_1926b,schrodinger_quantisierung_1926}.
In several aspects, this picture resembled Bohr's early theory of atomic orbits~\cite{Bohr1913OnTheConstitution}, which, on the one hand, was the spark of the quantum theory, and on the other hand, had been later discarded by the theory itself.

Since then, the Pauli master equation~\eqref{eq:Pauli-master} has been derived under various assumptions and with different techniques~\cite{BreuerPetruccione2002,Joos2003,schlosshauer_decoherence_2007}.
Notably, in the light of the decoherence theory for the quantum-to-classical transition developed since the 1970s~\cite{zeh_interpretation_1970,zeh_toward_1973,zurek_environment-induced_1982,zurek_decoherence_1991}, 
the Pauli master equation has been obtained for a quantum mechanical system coupled to the environment with appropriate assumptions~\cite{Joos2003,Joos1984}. The derivations were also discussed in close relation to the quantum Zeno effect~\cite{Joos2003,Joos1984,presilla_measurement_1996,FacchiPascazio2008Review}.

\begin{figure}[!tbh]
    \centering
    \includegraphics[width=\linewidth]{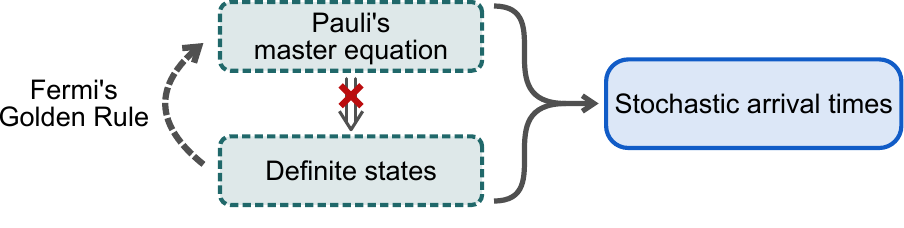}
    \caption{Schematic description \new{of the logical structure of the emergence of the classical stochasticity}.
    The decoherence theory for the quantum-to-classical transition indicates that the Pauli master equation does not imply the notion of system existing in definite states, leaving the picture of classical stochasticity incomplete.
    On the other hand, assuming that the system accommodates definite states, the Pauli master equation can be obtained by the application of the Fermi Golden Rule.
    Generally, the standard derivations of stochastic times assume not only the Pauli master equation, but also that system accommodates definite states at intermediate time points.}
    \label{figresults}
\end{figure}

However, the derivations of the Pauli master equation in the context of the decoherence theory for the quantum-to-classical transition are often regarded merely as a technical approximation of the open quantum dynamics, which follows more complicated quantum master equation (often also approximate) describing the evolution of the density operator~\cite{BreuerPetruccione2002,Joos2003,schlosshauer_decoherence_2007}. In particular, the founders of the decoherence theory themselves emphasised that even when the evolution of a quantum system can be well-approximated by eq.~\eqref{eq:Pauli-master}, one should not interpret the resulted density operator as `a proper mixture' of states distributed over $\{\ket{\mu}\}$~\cite[\new{p.~38}]{Joos2003}; see also Refs.~\cite{schlosshauer_decoherence_2007,zurek_decoherence_2003,schlosshauer_decoherence_2005,schlosshauer_elegance_2011,schlosshauer_quantum_2019,kiefer_quantum_2022}. Thus, this does not really complete the emergent classical stochasticity picture expected by Pauli, in which the quantum mechanical system accommodates \emph{definite states} and \emph{quantum jumps} take place stochastically. Let us emphasise that this picture is a crucial assumption behind Dirac's theory of Fermi's Golden Rule~\cite{Dirac1927EmissionAbsorption}, which is fundamental to Pauli's original derivation of his master equation, as well as to the discussions of various phenomena ranging from the quantum Zeno effect~\cite{Joos2003,Joos1984,presilla_measurement_1996,FacchiPascazio2008Review} to the theory of standard photon detectors in quantum optics~\cite{vogel2006quantum,WallsMilburn1994QuantumOptics}.
From the perspective of the decoherence theory for the quantum-to-classical transition~\cite{zeh_interpretation_1970,zeh_toward_1973,zurek_environment-induced_1982,zurek_decoherence_1991,schlosshauer_decoherence_2007,zurek_decoherence_2003,schlosshauer_decoherence_2005,schlosshauer_elegance_2011,schlosshauer_quantum_2019,kiefer_quantum_2022}, this assumption is, however, beyond what is conveyed by eq.~\eqref{eq:Pauli-master}; see Fig.~\ref{figresults}.  

\new{Here, we argue that the completion of Pauli's picture can be accomplished by considering the so-called ultradecoherence limit~\cite{Nguyen2025Ultradecoherence}}. The ultradecoherence limit refers to the particular regime where the decoherence takes place faster than any other relevant time scale~\cite{Nguyen2025Ultradecoherence}.
\new{We first show that the Pauli master equation~\eqref{eq:Pauli-master} can be obtained without explicitly assuming the system being in a definite state. This is different from the standard derivation using the Fermi Golden Rule~\cite{Pauli1928,Dirac1927EmissionAbsorption,Joos2003,Joos1984}, which assumes that the system exists relatively stably in definite states before computing the transition rates between them.}
\new{We then show that the requirement that the system exists in definite state can be supported by a different argument.}
\new{Indeed, as in this limit, the decoherence time scale is fast and classical processes couple to the system at a much slower time scale than that of the decoherence.}
In that case, the system can be considered to be in a definite state at all (coarser) time points when viewed relative to the relevant coupled classical processes~\cite{Nguyen2025Ultradecoherence}, which completes Pauli's picture of emergent classical stochasticity.

We then point out that stochastic times (such as persistent time, time of first arrivals) are uniquely defined once the system can be assumed to be in a definite state; see Fig.~\ref{figresults}.
This contrasts with the situation for quantum mechanical systems \new{with all possible superpositions allowed}, where there is no consensus on the definition of the time of arrivals~\cite{muga_time_2008,muga_time_2009}.
It is now important to emphasise that the unique definitions of such stochastic times are \emph{not} a consequence of the master equation~\eqref{eq:Pauli-master} alone. In \new{classical stochastic} processes, the possibility of conditioning probabilities of events on a particular \emph{definite} state of the system is crucial to the derivations of stochastic times~\cite{stirzaker_stochastic_2005}.
Such derivations break down if the system cannot be assumed to be in a definite state. This assumption is so natural in the classical theory of stochastic processes~\cite{stirzaker_stochastic_2005}, that it was never subject to question. 
For quantum systems, however, this is generally not the case; the validity of eq.~\eqref{eq:Pauli-master} is also not a sufficient condition~\cite{Joos2003,schlosshauer_decoherence_2007,zurek_decoherence_2003,schlosshauer_decoherence_2005,schlosshauer_elegance_2011,schlosshauer_quantum_2019,kiefer_quantum_2022}. 
The mentioned ultradecoherence limit is but a special case. 

In Section~\ref{sec:single-particle}, we systematically reconsider \new{some of} the logical details of the emergence of classical stochasticity. In particular, we consider the derivation of the Pauli master equation for single particles and how it is related to the stochastic time of arrivals. The novelty of our work lies in the interpretation and logical conditions required to endow a classical stochastic meaning to the dynamical equations, the latter being of course well-known for years.

In Section~\ref{sec:boson} and Section~\ref{sec:fermion}, we derive the master equations for fermions and bosons under the same assumptions. While these bosonic and fermionic master equations are also widely used in statistical physics, their derivations are less often discussed in detail in the literature. The derivation of the bosonic and fermionic master equations we present, on the other hand, allows us to analyse the quantum-to-classical transition of identical particles via decoherence. Indeed, following the above discussion, fermions and bosons subject to a fast decoherence process can be considered to be the realisation of classical identical particles. Notice that the concepts of classical identical particles were often formulated as a (hypothesised) starting basis for a quantised theory of identical particles~\cite{LeinaasMyrheim1977}. Assuming quantum mechanics as a fundamental starting point, our analysis concentrates on the reverse question of how the classical identical particles can actually emerge from quantum ones. 

\new{In Section~\ref{sec:transport}, using an example of one-dimensional transport, we show that concepts such as the Pauli exclusion principle of fermions or the non-equilibrium Bose--Einstein condensation of bosons can manifest themselves for classical identical particles at ultradecoherence.
In addition, we show that such fermions and bosons at ultradecoherence also have well-defined time of arrivals~\cite{muga_time_2008,muga_time_2009}.}

\section{Single particle stochastic dynamics}
\label{sec:single-particle}

\subsection{Single particle master equation}
Consider a quantum mechanical system in coupling with the environment as typically considered in the decoherence theory~\cite{BreuerPetruccione2002,Joos2003,schlosshauer_decoherence_2007}. Without much detail of this coupling, it is assumed that the system follows a Lindblad master equation with a fast dephasing in a basis $\{\ket{ \mu }\}$~\cite{BreuerPetruccione2002,Joos2003,schlosshauer_decoherence_2007}. Following the decoherence literature~\cite{zurek_decoherence_2003,schlosshauer_decoherence_2005,schlosshauer_decoherence_2007}, the basis $\{\ket{ \mu }\}$ is referred to as the \emph{preferred basis}. 
For concreteness, one can imagine a particle in a lattice with positions corresponding to states of the preferred basis. 
For the derivation of the Pauli master equation, the assumption that the particle is in a definite state is not made at the outset.

The free Hamiltonian of the system is deliberately decomposed into the unperturbed contribution $H_D$, which is diagonal in the preferred basis with eigenvalues $\{\Omega_\mu\}$, and the interaction term $V$ having no diagonal elements, $V_{\mu\mu}=0$. In general, the diagonal terms $\{\Omega_\mu\}$ represent intrinsic energies of the states $\ket{\mu}$, and the off-diagonal terms $V_{\mu \nu}$ of the interaction $V$ are responsible for the quantum tunneling of the system between different states.
We are to see that, under strong decoherence, this quantum mechanical tunnelling effect degrades to classical stochastic transitions.

Formally, the Lindblad master equation for the system is assumed to be~\cite{BreuerPetruccione2002,Joos2003,schlosshauer_decoherence_2007}
\begin{equation}
    \frac{d \rho}{d t} = -i[H_D +V,\rho] + \sum_{ \mu } \gamma_{\mu} \left( L_{\mu} \rho L_{\mu} - \frac{1}{2} \{L_{\mu},\rho\} \right),
    \label{eq:simple-lindblad}
\end{equation}
where $L_{\mu}=\ketbra{\mu}{\mu}$ are the Lindblad operators and $\gamma_\mu$ are the dephasing rates. We use the natural quantum mechanical unit throughout this article, so that $\hbar=1$ in all expressions. Here we are particularly interested in the so-called ultradecoherence limit~\cite{Nguyen2025Ultradecoherence}, where the dephasing dynamics is much faster than any other relevant time scales, in particular, $\gamma_{\mu} \gg \abs{V_{\nu \lambda}}$.
The following analysis bears some similarity to that of the model of the measurement process in Ref.~\cite{Nguyen2025Ultradecoherence}, which consists of a quantum mechanical system coupled to a measurement device embedded in an environment. Involving however only one system coupled to the environment, the analysis here is somewhat simpler and will be presented in more detail for pedagogical purposes.

One starts with explicitly expressing eq.~\eqref{eq:simple-lindblad} in the preferred basis, which gives
\begin{equation}
    \frac{d \rho_{\mu \nu}}{d t} = -i\sum_{\lambda} \left(V_{\mu \lambda} \rho_{\lambda \nu} - \rho_{\mu \lambda} V_{\lambda \nu} \right) - (i \Omega_{\mu \nu}+\gamma_{\mu \nu}) \rho_{\mu \nu},
    \label{eq:V-master-full}
\end{equation}
where one defines $\Omega_{\mu \nu} = \Omega_{\mu} - \Omega_{\nu}$ and
\begin{equation}
    \gamma_{\mu \nu}= \left\{ \begin{array}{cc} \frac{1}{2} (\gamma_{\mu} + \gamma_{\nu}) \mbox{ if $\mu \ne \nu$}, \\ 0 \mbox{ if $\mu = \nu$}. \end{array}\right.
\end{equation}

\new{Given the preferred basis defined by the coupling with the environment}, recall that the off-diagonal terms of the density operators $\rho_{\mu \nu}$ for $\mu \ne \nu$ are called \emph{coherences}~\cite{BreuerPetruccione2002,Joos2003,schlosshauer_decoherence_2007}.
Assuming the decoherence takes place at dominating rates $\gamma_\mu$,
one can carry out an adiabatic elimination of the coherences. To this end, for $\mu \ne \nu$, one implicitly solves $\rho_{\mu \nu}$ from eq.~\eqref{eq:V-master-full} by the usual method of Green's function as
\begin{widetext}
\begin{equation}
    \rho_{\mu \nu} (t) = -i \int_{0}^{t} \mathrm{d} \tau e^{-(i\Omega_{\mu \nu}+\gamma_{\mu \nu}) \tau} \sum_{\lambda} \left[ V_{\mu \lambda} \rho_{\lambda \nu} (t-\tau) - \rho_{\mu \lambda} (t-\tau) V_{\lambda \nu} \right] \quad \red{(\mu \neq \nu)},
\end{equation}
\end{widetext}
where it is assumed that the coherences are not present in the initial state.
The first term under the summation can be written as
\begin{multline}
    V_{\mu \lambda} \int_{0}^{t} \mathrm{d} \tau e^{-(\gamma_{\mu \nu}+ i\Omega_{\mu \nu}) \tau} \rho_{\lambda \nu} (t-\tau) = \\ \quad V_{\mu \lambda} \int_{0}^{t/\gamma_{\mu \nu}} \mathrm{d} \tau' e^{- (1+i \frac{\Omega_{\mu \nu}}{\gamma_{\mu \nu}})\tau'} \frac{\rho_{\lambda \nu} (t-\frac{\tau'}{\gamma_{\mu \nu}})}{\gamma_{\mu \nu}}.
\end{multline}
Assuming that $\gamma_{\mu \nu}$ is large in comparison to the rate at which the density operator varies, one can approximate $\rho_{ \lambda \nu} (t-\frac{\tau'}{\gamma_{\mu \nu}}) \approx \rho_{\lambda\nu} (t)$ and the integral limit can be approximated as $t/\gamma_{\mu \nu} \approx \infty$. This is often known in the theory of open quantum systems as the Markovian approximation~\cite{BreuerPetruccione2002,Joos2003,schlosshauer_decoherence_2007}. Within this approximation, one has
\begin{equation}
    V_{\mu \lambda} \int_{0}^{t} \mathrm{d} \tau e^{-(\gamma_{\mu \nu}+ i\Omega_{\mu \nu}) \tau}  \rho_{\lambda\nu} (t-\tau)  \approx 
    \frac{V_{\mu \lambda}}{\gamma_{\mu \nu} + i \Omega_{\mu \nu }} \rho_{\lambda \nu }.
\end{equation}

Similarly,
\begin{equation}
     \int_{0}^{t} \mathrm{d} \tau e^{-(\gamma_{\mu \nu }+ i\Omega_{\mu \nu }) \tau}  \rho_{\mu\lambda} (t-\tau) V_{\lambda \nu}  \approx
    \rho_{\mu\lambda} \frac{V_{ \lambda \nu }}{\gamma_{\mu \nu } + i \Omega_{\mu \nu}} .
\end{equation}

Eventually, one thus obtains, \red{still for $\mu \ne \nu$},
\begin{equation}
    \rho_{\mu \nu} = -i   \sum_{\lambda} \left[\frac{V_{\mu \lambda}}{\gamma_{\mu \nu} + i \Omega_{\mu \nu }} \rho_{\lambda \nu} -  \rho_{\mu \lambda} \frac{V_{\lambda \nu}}{\gamma_{\mu \nu} + i \Omega_{\mu \nu}}\right].
    \label{eq:coherence-solved}
\end{equation}
Equation~\eqref{eq:coherence-solved} allows one to solve for the coherence terms of the density operator $\rho_{\mu \nu}$ in terms of the diagonal elements. In particular, one can eliminate all coherence terms from the Lindblad eq.~\eqref{eq:V-master-full} to obtain an equation of evolution for the diagonal elements only.

The solution of eq.~\eqref{eq:coherence-solved} can be greatly simplified in the limit where $\abs{\gamma_{\mu \nu} + i \Omega_{\mu \nu}}\gg \abs{V_{\lambda \kappa}}$ for all $\mu,\nu,\lambda,\kappa$ \red{with $\mu\neq\nu$}. 
Considering that eq.~\eqref{eq:coherence-solved} is solved by iteration starting with the trial solution $\rho_{\mu \nu} = 0$ for $\mu \ne \nu$ until convergence.
One sees that each iteration gives a correction to the solution of the increasing order in $\mathcal{O} (1/\abs{\gamma_{\mu \nu} + i \Omega_{\mu \nu}})$.
In particular, to the first order of $\mathcal{O} (1/\abs{\gamma_{\mu \nu} + i \Omega_{\mu \nu}})$, only the diagonal terms of the density operator contribute to the summation on the right hand side of eq.~\eqref{eq:coherence-solved}, and one obtains a rather simple approximate solution
\begin{equation}
    \rho_{\mu \nu } \approx   \frac{-i V_{\mu \nu}  }{\gamma_{\mu \nu} + i \Omega_{\mu \nu}}   \left( \rho_{\nu \nu} -  \rho_{\mu \mu} \right) \quad \red{(\mu \neq \nu)}.
    \label{eq:adiabatic-coherence}
\end{equation}

To obtain the dynamical equation for the diagonal terms of the density operator only, one substitutes eq.~\eqref{eq:adiabatic-coherence} into eq.~\eqref{eq:V-master-full} and arrives at 
\begin{equation}
    \frac{\mathrm{d} P_{\mu}}{\mathrm{d} t} = \sum_{\nu \ne \mu} W_{\mu \nu} (P_{\nu} - P_{\mu}),
    \label{eq:master-classical}
\end{equation}
where $P_{\mu}= \rho_{\mu \mu}$ and
\begin{equation}
   W_{\mu \nu}= \frac{2\gamma_{\mu \nu} \abs{V_{\mu \nu}}^2}{\gamma_{\mu \nu}^2 + \Omega_{\mu \nu}^2}. 
   \label{eq:V-master-transition}
\end{equation}

Equation~\eqref{eq:master-classical} resembles \new{a master equation in the classical theory of stochastic processes} describing the evolution of the probability for the system to be in states $\mu$, $P_{\mu}= \rho_{\mu \mu}$, with time-independent symmetric transition rate $W_{\mu  \nu}$ given by eq.~\eqref{eq:V-master-transition}.

To be consistent with the adiabatic elimination, one requires that $W_{\mu \nu} \ll \gamma_{\lambda}$. Assuming the decoherence rates $\gamma_{\lambda}$ for different states are of the same order, this amounts to
\begin{equation}
    \abs{V_{\mu \nu}}^2 \ll \gamma_{\mu \nu}^2 + \Omega_{\mu \nu}^2.
\end{equation}
Observe that this condition can also be satisfied even when the decoherence rates $\gamma_{\mu \nu}$ are relatively small, but the energy separations $\Omega_{\mu \nu}$ are large in comparison to the tunnelling elements, $\abs{V_{\mu \nu}}^2 \ll \Omega_{\mu \nu}^2$. \new{In this case, the large energy separations $\Omega_{\mu \nu}$ cause the blocks of the density operator to evolve slowly and the adiabatic approximation can still be expected to hold.}

The transition rate $W_{\mu \nu}$ strongly resembles the classic Fermi Golden Rule~\cite{Dirac1927EmissionAbsorption}, with ${\gamma_{\mu \nu}}/[\pi ({\gamma_{\mu \nu}^2 + \Omega_{\mu \nu}^2})]$ playing the role of the density of states at the scattered levels. This is intuitively understandable from the decoherence theory, as the decoherence due to coupling to the environment is also the cause of the spreading of the energy level $\Omega_\mu$ of system to a lorentzian form with width $\gamma_{\mu}$. Considering the limit $\gamma_{\mu \nu} \ll \abs{\Omega_{\mu \nu}}$, one can approximate the lorentzian density of states by a Dirac $\delta$-function, ${\gamma_{\mu \nu}}/[\pi ({\gamma_{\mu \nu}^2 + \Omega_{\mu \nu}^2})] \approx \delta (\Omega_{\mu \nu})$ and recover the familiar form of the Fermi Golden Rule,
\begin{equation}
   W_{\mu \nu} \approx {2 \pi \abs{V_{\mu \nu}}^2} \delta(\Omega_{\mu \nu}).
\end{equation}

Now it is important to remark that the time-dependent perturbation theory was not explicitly invoked as in the standard derivation of the Fermi Golden Rule~\cite{Dirac1927EmissionAbsorption}. The standard derivation assumes that the system exists relatively stably in definite states \emph{before} computing the transition rates between them. This was not among the assumptions leading to eq.~\eqref{eq:master-classical} in the above derivation. 

Considering the opposite limit $\gamma_{\mu \nu} \gg \Omega_{\mu \nu}$, one obtains
\begin{equation}
    W_{\mu \nu} \approx \frac{2 \abs{V_{\mu \nu}}^2}{\gamma_{\mu \nu}}.
    \label{eq:zeno}
\end{equation}
This expression is a well-known manifestation of the quantum Zeno effect~\cite{Joos2003,Joos1984,presilla_measurement_1996,FacchiPascazio2008Review}.
\new{Originally, the quantum Zeno effect refers to the fact that measurements continuously made on the system freeze its dynamics~\cite{presilla_measurement_1996,FacchiPascazio2008Review}. One way to understand the quantum Zeno effect is to see that such measurements continuously project the state of the system back onto its current state. A more dynamical way of understanding the effect is to consider a system coupled to the surrounding environment, which then follows an open quantum dynamics~\cite{Joos2003,Joos1984}. Results in the spirit of eq.~\eqref{eq:zeno} were obtained: the dynamics of quantum tunnelling between energy levels is suppressed by a factor of $1/\gamma_{\mu \nu}$ characterising the coupling strength to the environment~\cite{Joos2003,Joos1984,presilla_measurement_1996,FacchiPascazio2008Review}.} 
In particular, in the limit $\gamma_{\mu \nu} \to \infty$, the system dynamics is frozen.
In the ultradecoherence regime we considered here, $\gamma_{\mu \nu}$ is assumed to be large, but finite.

It is also remarkable that $W_{\mu \nu}$ is symmetric, $W_{\mu \nu} = W_{ \nu \mu}$. It implies that a transition from a higher level to a lower energy level is taking place with the same rate as the reverse transition. As a result, the detailed balance is satisfied with a constant distribution $P_{\mu}= \mathrm{const}$~\cite{stirzaker_stochastic_2005}. Therefore, the system can be considered as being effectively at infinite temperature. This is consistent with the general literature of open quantum systems, where it is known that pure dephasing pumps energy into the system~\cite{Joos2003}.

\subsection{Persistent time and time of first arrivals}
\label{sec:time}

Following the perspective of the decoherence theory for the quantum-to-classical transition~\cite{zeh_interpretation_1970,zeh_toward_1973,zurek_environment-induced_1982,zurek_decoherence_1991,schlosshauer_decoherence_2007,zurek_decoherence_2003,schlosshauer_decoherence_2005,schlosshauer_elegance_2011,schlosshauer_quantum_2019,kiefer_quantum_2022}, one can argue that the master equation~\eqref{eq:master-classical} does not imply that the system can be assumed to be in a definite state. For the ultradecoherence limit, however, the classical processes couple to the system at a much slower time scale than that of the decoherence time scale~\cite{Nguyen2025Ultradecoherence}. 
Thus, the system can be considered to be in a definite state at all (coarser) time points when considered with respect to the relevant coupled classical processes~\cite{Nguyen2025Ultradecoherence}. 
This aligns well with and is often implicit in the discussion of the quantum Zeno effect~\cite{Joos2003,Joos1984,presilla_measurement_1996,FacchiPascazio2008Review}.
In fact, one can interpret that it is the mechanism of the quantum Zeno effect that allows for the stable existence of the system (frozen) in a definite state. Before the regime of the full quantum Zeno effect ({$\gamma_{\mu \nu} \to \infty$}), the quantum tunnelling dynamics, however, remains and manifests itself as classical jumps between these different states. 

A system that exists in a definite state has an interesting independent consequence: various stochastic times are well-defined.
Indeed, the standard derivations of the stochastic times in the classical theory of stochastic processes~\cite{stirzaker_stochastic_2005} are crucially based on the probabilities \emph{conditioned} on the fact that the system exists in a particular state at certain time. 
The computation of the time of first arrivals is somewhat technical for the purpose of illustration; see, e.g., Ref.~\cite{stirzaker_stochastic_2005} and also Appendix~\ref{sec:app-arrivals}. 
Sufficiently to highlight the conceptual argument without much technicality, we consider here the computation of the persistent time at a state. The discussion of the time of first arrivals by means of Monte-Carlo simulation is also considered in Section~\ref{sec:transport}. 

Let the system be in a state $\ket{\mu}$. One asks: when does the system first leave the state $\ket{\mu}$? This is referred to as the persistent time $T_{\mu}$ at state $\ket{\mu}$.
\new{The persistent time is thus also the arrival time at the set of states other than $\ket{\mu}$.}
Clearly $T_{\mu}$ is a random variable, and we are to compute the tail distribution $P(T_{\mu}>t)$, which is the probability that the system remains in the state $\ket{\mu}$ until time $t$. 

Assuming that the system has remained in state $\ket{\mu}$ until time $t$, we consider the system after an infinitesimal time $\Delta t$. Conditioned on the fact that the system is at state $\ket{\mu}$ until time $t$, the probability that no transition takes place between $t$ and $t+\Delta t$ is given by $1-\sum_{\nu \ne \mu } W_{\nu \mu} \Delta t$. The probability that the system remains at state $\ket{\mu}$ until $t+\Delta t$, $P(T_{\mu}>t+\Delta t)$, is then 
\begin{equation}
    P(T_{\mu}>t+\Delta t) = (1-\sum_{\nu \ne \mu } W_{\nu \mu} \Delta t) P(T_{\mu}>t).
\end{equation}
Taking the limit $\Delta t \to 0$, one obtains a differential equation
\begin{equation}
    \frac{d}{dt} P(T_{\mu}>t)= -\left(\sum_{\nu \ne \mu } W_{\nu \mu}\right) P(T_{\mu}>t),
\end{equation}
which can be solved with the initial condition $P(T_{\mu}>0) = 1$ by
\begin{equation}
    P(T_{\mu}>t) = \exp\left\{-t\sum_{\nu \ne \mu } W_{\nu \mu}\right\}.
    \label{eq:first-transition-time}
\end{equation}

While the above derivation is standard and straightforward, notice that it crucially relies on the ability of conditioning events on the fact that `the system is at state $\ket{\mu}$ until time $t$.' Without being able to assume that the system is in a definite state, it is unclear how to justify such calculation. This remains so even when a master equation of the form eq.~\eqref{eq:Pauli-master} or eq.~\eqref{eq:master-classical} is valid~\cite{zeh_interpretation_1970,zeh_toward_1973,zurek_environment-induced_1982,zurek_decoherence_1991,schlosshauer_decoherence_2007,zurek_decoherence_2003,schlosshauer_decoherence_2005,schlosshauer_elegance_2011,schlosshauer_quantum_2019,kiefer_quantum_2022}; see also Appendix~\ref{sec:app-decoupled-diagonals}. 
As a result, there are several inequivalent approaches to the time of arrivals in quantum mechanics~\cite{muga_time_2008,muga_time_2009}. 

We have discussed the computation of the persistent time. While technically more demanding, the computation of the time of first arrivals also similarly requires conditioning the system on a certain state; see Appendix~\ref{sec:app-arrivals}. In Section~\ref{sec:transport}, the arrival times are derived from Monte-Carlo simulations. Explicitly, the Monte-Carlo simulations themselves assume that the system is in a definite state at a given time. 

\section{Classical stochastic dynamics of bosons}
\label{sec:boson}
\subsection{Bosonic master equation}

Let us now consider a system of bosonic particles distributed over $N$ modes.
For concreteness, one can regard the modes as `positions' of the particles in a lattice. 
The Hamiltonian of the system can be written in the second-quantised form as
\begin{equation}
    H = \sum_{\mu} \Omega_{\mu} a^{\dagger}_\mu a_\mu+ \sum_{\mu \ne \nu} V_{\mu \nu} a^{\dagger}_\mu a_\nu,
\end{equation}
where the bosonic annihilators $a_\mu$ and creators $a_\mu^\dagger$ satisfy
\begin{equation}
    \mbox{$[a_{\mu},a_{\nu}^\dagger]=\delta_{\mu\nu}$, $[a_{\mu},a_{\nu}]=0$, $[a_{\mu}^{\dagger},a_{\nu}^\dagger]=0$. }
\end{equation}
In similarity to the investigation of the dynamics of a single particle, we have also deliberately separated the diagonal parts containing $\Omega_{\mu}$ and the off-diagonal part containing $V_{\mu \nu}$ of the Hamiltonian for convenience. 

It is then assumed that the system is subject to a fast dephasing decoherence in the positions $\mu$.  
More precisely, the equation of motion is assumed to be of the Lindblad form,
\begin{equation}
    \frac{d \rho }{d t} = -i [H,\rho] + \sum_{\mu} \gamma_{\mu} \left( n_\mu \rho n_\mu - \frac{1}{2} \{n_\mu^2, \rho \} \right),
    \label{eq:lindblad-boson-operator}
\end{equation}
where $n_\mu = a^{\dagger}_\mu a_\mu$ are the occupation operators and $\gamma_\mu$ are the decoherence rates. This equation should be regarded as the bosonic second-quantised analogue of the single particle master equation~\eqref{eq:simple-lindblad}.

Under this decoherence process, it is natural to express eq.~\eqref{eq:lindblad-boson-operator} in the Fock basis. For such a system of $N$ bosonic modes, a Fock state is denoted by an $N$-dimensional non-negative integer vector $\ket{\vec{m}} = \ket{\{m_{\mu}\}}$, which is explicitly constructed as
\begin{equation}
    \ket{\vec{m}} = \prod_{\mu=1}^{N}\frac{(a_\mu^\dagger)^{m_\mu}}{\sqrt{m_\mu!}}\ket{0},
\end{equation}
where $\ket{0}$ represents the vacuum state. 
The density matrix can be decomposed in this basis as
\begin{equation}
    \rho = \sum_{\vec{m}, \vec{n}}\rho_{\vec{m},\vec{n}} \ket{\vec{m}}\bra{\vec{n}}.
\end{equation}

Expressing in the Fock basis, the Lindblad master equation~\eqref{eq:lindblad-boson-operator} for the matrix elements becomes
\begin{widetext}
\begin{align}
    \frac{d}{dt}\rho_{\vec{m}, \vec{n}} = & -i\left[\rho_{\vec{m},\vec{n}} \sum_\mu \Omega_{\mu}(m_\mu-n_\mu)+\sum_{\mu \ne \nu}V_{\mu \nu }\left( \sqrt{(m_\nu+1)m_\mu}\,\rho_{\vec{m}-\vec{e}_\mu+\vec{e}_\nu,\vec{n}}-\sqrt{(n_\mu+1)n_\nu}\,\rho_{\vec{m},\vec{n}+\vec{e}_\mu-\vec{e}_{\nu}}\right)\right] \notag\\
    &-\frac{1}{2}\rho_{\vec{m}, \vec{n}}\sum_{\nu}\gamma_{\nu} (m_\nu-n_\nu)^2,
    \label{eq:lindblad-boson-element}
\end{align}
\end{widetext}
where $\vec{e}_{\mu}$ denotes the unit vector in the $\mu$-th direction, with components $(\vec{e}_{\mu})_{\nu} = \delta_{\mu \nu}$. Notice that the prefactors $\sqrt{(m_\nu+1)m_\mu}$ and $\sqrt{(n_\mu+1)n_\nu}$ ensure that states with negative occupation numbers do not actually contribute in eq.~\eqref{eq:lindblad-boson-element}. 

We again concentrate on the ultradecoherence limit, where the decoherence rates $\gamma_\mu$ are dominating. Observe that if $\vec{m}\neq\vec{n}$, the presence of the second term in \eqref{eq:lindblad-boson-element} again allows one to carry out an adiabatic elimination in similarity to Section~\ref{sec:single-particle}. 
This then leads to, in place of eq.~\eqref{eq:coherence-solved},
\begin{widetext}
\begin{align}
\rho_{\vec{m},\vec{n}} \approx &  \frac{- i \sum_{ \mu \neq \nu}  V_{\mu \nu}\left( \sqrt{(m_\nu+1)m_\mu}\cdot\rho_{\vec{m}-\vec{e}_\mu +\vec{e}_{\nu},\vec{n}}-\sqrt{(n_\mu+1)n_\nu}\cdot\rho_{\vec{m},\vec{n}+\vec{e}_{\mu}-\vec{e}_{\nu}}\right)}{\frac{1}{2}\sum_\lambda \gamma_\lambda(m_\lambda-n_\lambda)^2+ i\sum_\mu \Omega_{\mu} (m_\mu-n_\mu)} \quad \red{(\vec{m} \neq \vec{n})}.
    \label{eq:bosonic-off-diagonal}
\end{align}
\end{widetext}
One then proceeds to obtain a simpler approximation as an analogue of eq.~\eqref{eq:adiabatic-coherence} as follows. On the basis of eq.~\eqref{eq:bosonic-off-diagonal}, one assumes that the diagonal $\rho_{\vec{m}, \vec{m}}$ is of the order $\mathcal{O}(1)$, and the coherence (off-diagonal) elements $\rho_{\vec{m},\vec{n}}$ ($\vec{m} \ne \vec{n}$) are of the order $\mathcal{O}(\gamma_{\lambda}^{-1})$. To this order $\mathcal{O}(\gamma_{\lambda}^{-1})$, however, the coherences themselves can be ignored in the right-hand side of eq.~\eqref{eq:bosonic-off-diagonal}. 
As a result, one finds that for any pair $(\vec{m}, \vec{n})$ \red{with $\vec{m} \ne \vec{n}$} where there exist no indices $(\mu, \nu)$ such that $\vec{m} - \vec{e}_\mu + \vec{e}_{\nu} = \vec{n}$, one can approximate $\rho_{\vec{m}, \vec{n}} \approx 0$ to the order $\mathcal{O}(\gamma_{\lambda}^{-1})$.
The non-vanishing coherence elements are thus when $\vec{m}-\vec{e}_{\mu}+\vec{e}_{\nu}=\vec{n}$.
One can define $\pmb{s}_{\mu \nu}=\vec{e}_{\mu}-\vec{e}_{\nu}$, which describes an elementary process of moving a particle initially at position $\nu$ to position $\mu$. The non-vanishing coherence elements are then
\begin{align}
\rho_{\vec{m},\vec{m}-\pmb{s}_{\mu \nu}}&\approx  \frac{ -i V_{\mu \nu} \sqrt{m_\mu(m_\nu+1)} }{\gamma_{\mu \nu}+i\Omega_{\mu\nu}} \left( \rho_{\vec{m}-\vec{s}_{\mu\nu},\vec{m}-\vec{s}_{\mu\nu}}- \rho_{\vec{m},\vec{m}}\right),
\label{eq:bosonic-off-diagonal_term-simplified}
\end{align}
where $\Omega_{\mu \nu} = \Omega_{\mu} - \Omega_{\nu}$, $\gamma_{\mu \nu} = (\gamma_{\mu}+\gamma_{\nu})/2$ if $\mu \ne \nu$ and $0$ otherwise, which are identical with the definitions for the case of a single particle considered in Section~\ref{sec:single-particle}. 

Inserting the adiabatic approximation for the coherences in eq.~\eqref{eq:bosonic-off-diagonal_term-simplified} into the Lindblad master equation~\eqref{eq:lindblad-boson-element} for $\vec{m}=\vec{n}$, one obtains 
\begin{align}
    \frac{d P_{\vec{m}}}{dt} = & \sum_{\mu \neq \nu}  W_{\mu \nu} (m_\mu+1) m_\nu  \left(P_{\vec{m}+\vec{s}_{\mu \nu}} - P_{\vec{m}}\right),
    \label{eq:classical-master-boson}
\end{align}
where $P_{\pmb{m}}=\rho_{\vec{m},\vec{m}}$ and the single particle transition rate $W_{\mu \nu}$ is identical with the definition~\eqref{eq:V-master-transition}. In comparison with the single particle master equation~\eqref{eq:master-classical}, one sees that the transition rate of the scattering a particle in mode $\nu$ into mode $\mu$ acquires the familiar bosonic stimulation factor of $(m_\mu+1) m_\nu$, known since Einstein's seminal work~\cite{Einstein1917}.

Notice that the transition rate in the master equation~\eqref{eq:classical-master-boson} is symmetric, implying that the equilibrium distribution is uniform in Fock space. The system is thus again effectively at infinite temperature, which also aligns with the discussion in Section~\ref{sec:single-particle}.

\subsection{Loss and pump}
In many situations of interest, the system might constantly suffer from loss of particles. On the other hand, new particles  can also be pumped into the system.
To model these effects, we consider the modification of the Lindblad master equation~\eqref{eq:lindblad-boson-operator} to explicitly contain terms corresponding to loss and pump,
\begin{align}
    \frac{d \rho }{d t} =& -i [H,\rho] + \sum_{\mu} \gamma_{\mu} \left( n_\mu \rho n_\mu - \frac{1}{2} \{n_\mu^2, \rho \} \right) \nonumber \\
    & \quad  + \sum_{\nu} \theta_{\nu} \left( a_{\nu} \rho a_{\nu}^{\dagger} - \frac{1}{2}\{a_\nu^\dagger a_\nu, \rho \} \right) \nonumber \\
    & \quad + \sum_{\nu} \eta_{\nu} \left(a_{\nu}^\dagger \rho a_{\nu} - \frac{1}{2}\{a_\nu a_\nu^\dagger, \rho \} \right),
    \label{eq:lindblad-boson-operator-pump-and-loss}
\end{align}
where $\theta_{\nu}$ is the loss rate and $\eta_{\nu}$ is the pump rate at mode $\nu$.
These loss and pump terms in this form are familiar in the literature of quantum optics~\cite{WallsMilburn1994QuantumOptics}.
Notice that even in the presence of loss and pump, we still assume that the dephasing rate $\gamma_{\mu}$ is the dominant one, $\gamma_{\mu} \gg \eta_{\nu}, \theta_{\nu}$. 

The evolution of the matrix elements of the density operator in the Fock basis, in place of eq.~\eqref{eq:lindblad-boson-element}, now reads:
\begin{widetext}
\begin{align}
    \frac{d\rho_{\vec{m}, \vec{n}}}{dt} = & -i\left[\rho_{\vec{m},\vec{n}} \sum_\mu \Omega_{\mu}(m_\mu-n_\mu)+\sum_{\mu \ne \nu}V_{\mu \nu }\left( \sqrt{(m_\nu+1)m_\mu}\,\rho_{\vec{m}-\vec{e}_\mu+\vec{e}_\nu,\vec{n}}-\sqrt{(n_\mu+1)n_\nu}\,\rho_{\vec{m},\vec{n}+\vec{e}_\mu-\vec{e}_{\nu}}\right)\right] \notag\\
    &-\frac{1}{2}\rho_{\vec{m}, \vec{n}}\sum_{\nu}\gamma_{\nu} (m_\nu-n_\nu)^2 \nonumber \\
    &+\sum_{\nu} \theta_{\nu} \sqrt{(m_{\nu}+1)(n_{\nu}+1)} 
\rho_{\vec{m}+\vec{e}_{\nu},\vec{n}+\vec{e}_{\nu}} - \frac{1}{2} \sum_{\nu} \theta_{\nu} (m_{\nu}+n_{\nu}) \rho_{\vec{m},\vec{n}} \nonumber \\
    &+\sum_{\nu} \eta_{\nu} \sqrt{m_{\nu}n_{\nu}} 
\rho_{\vec{m}-\vec{e}_{\nu},\vec{n}-\vec{e}_{\nu}} - \frac{1}{2} \sum_{\nu} \eta_{\nu} (2+m_{\nu}+n_{\nu}) \rho_{\vec{m},\vec{n}}.
    \label{eq:master_equation_pump_loss_boson}
\end{align}
\end{widetext}
To proceed, we assume again that the density operators are dominated by diagonal elements $\rho_{\vec{m},\vec{m}}$ of the order $\mathcal{O}(1)$, the coherences $\rho_{\vec{m},\vec{n}}$ for $\vec{m} \ne \vec{n}$ are small of the order of $\mathcal{O}(\gamma_{\nu}^{-1})$. 
We again assume that $\gamma_{\nu}$ are the dominating rates, $\gamma_{\lambda} \gg \eta_{\nu}, \theta_{\nu}$, such that the pump and loss in eq.~\eqref{eq:master_equation_pump_loss_boson} can also be ignored for the case $\vec{m} \ne \vec{n}$. 
In the end, one obtains again eq.~\eqref{eq:bosonic-off-diagonal} and eq.~\eqref{eq:bosonic-off-diagonal_term-simplified} for the coherences $\vec{m} \ne \vec{n}$ to the leading order contribution $\mathcal{O}(\gamma_{\nu}^{-1})$. Inserting the eq.~\eqref{eq:bosonic-off-diagonal_term-simplified} into eq.~\eqref{eq:master_equation_pump_loss_boson} for $\vec{m} = \vec{n}$, one obtains 
\begin{align}
    \frac{d P_{\vec{m}}}{dt} = & \sum_{\mu \neq \nu}  W_{\mu \nu} (m_\mu+1) m_\nu  \left(P_{\vec{m}+\vec{s}_{\mu \nu}} - P_{\vec{m}}\right) \nonumber    \\
    &\quad + \sum_\nu \theta_\nu \left[ (m_\nu+1)  P_{\vec{m}+\vec{e}_\nu} - m_\nu  P_{\vec{m}} \right] \nonumber \\
    &\quad + \sum_\nu \eta_\nu \left[ m_\nu P_{\vec{m}-\vec{e}_\nu} - (1+m_\nu)  P_{\vec{m}} \right],
    \label{eq:dynamic_with_creation_and_loss}
\end{align}
with $W_{\mu \nu}$ and $P_{\vec{m}}$ defined as in eq.~\eqref{eq:classical-master-boson}. 
In comparison to eq.~\eqref{eq:classical-master-boson}, the master equation~\eqref{eq:dynamic_with_creation_and_loss} gains new terms due to loss and pump as expected.

Again, it is important to emphasise that neither the derivation of equation~\eqref{eq:classical-master-boson} nor that of equation~\eqref{eq:dynamic_with_creation_and_loss} explicitly requires the assumption that the system is in a definite Fock state. 
As we mentioned, it is important that the Fermi Golden Rule was also not explicitly invoked.
Specifically for the ultradecoherence limit, however, the system can be assumed to be indeed in a definite state as an independent justified assumption, as discussed in Section~\ref{sec:time}. The consequence of this assumption for the computation of the stochastic time of first arrivals will be discussed in Section~\ref{sec:transport}.

\section{Classical stochastic dynamics of fermions}
\label{sec:fermion}

The derivation of the master equation for a fermionic system can proceed similarly to that for bosonic systems. It is, however, important to keep in mind the anticommutation of fermionic operators. As a particular consequence, the fermionic occupation number vector $\vec{m}$ now has only binary values ($0$ and $1$) for its components. Here, we sketch the main steps of the derivation.

One starts with the general Lindblad master equation with pump and loss for fermions in second-quantisation as 
\begin{align}
    \frac{d \rho }{d t} =& -i [H,\rho] + \sum_{\mu} \gamma_{\mu} \left( n_\mu \rho n_\mu - \frac{1}{2} \{n_\mu^2, \rho \} \right) \nonumber \\
    & \quad  + \sum_{\nu} \theta_{\nu} \left( b_{\nu} \rho b_{\nu}^{\dagger} - \frac{1}{2}\{b_\nu^\dagger b_\nu, \rho \} \right) \nonumber \\
    & \quad + \sum_{\nu} \eta_{\nu} \left(b_{\nu}^\dagger \rho b_{\nu} - \frac{1}{2}\{b_\nu b_\nu^\dagger, \rho \} \right), 
    \label{eq:fermion-lindblad-operator}
\end{align}
where the fermionic annihilators $b_{\mu}$'s and creators  $b^{\dagger}_{\nu}$'s follow the fermionic anticommutation relations,
\begin{equation}
    \{b^\dagger_{\mu},b_{\nu}\} = \delta_{\mu \nu}, \{b_{\mu},b_{\nu}\} = 0 , \{b^\dagger_{\mu},b^\dagger_{\nu}\} = 0.
\end{equation}
Recall that from the fermionic anticommutation relations, it then follows directly
\begin{align}
    b_{\mu} \ket{n_{\mu}} &= n_{\mu} \ket{1-n_{\mu}}= n_{\mu}  \ket{n_{\mu}-1}, \\
    b_{\mu}^{\dagger} \ket{n_{\mu}} &= (1-n_{\mu}) \ket{1-n_{\mu}} = (1-n_{\mu}) \ket{n_{\mu}+1},
\end{align}
with the convention that $n_{\mu} \in \{0,1\}$ and $n_{\mu}  \ket{n_{\mu}-1}=0$ for $n_{\mu}=0$ and $(1-n_{\mu}) \ket{n_{\mu}+1}=0$ for $n_{\mu}=1$ so that states of occupation numbers other than $0$ and $1$ do not actually arise.

For $N$ fermionic modes, a Fock state is denoted by an $N$-dimensional binary integer vector $\ket{\vec{m}} = \ket{\{m_{\mu}\}}$, whose components assume values of $0$ or $1$. Explicitly, the Fock state $\ket{\vec{m}}$ is  given as
\begin{equation}
    \ket{\vec{m}} = \prod_{\mu=1}^{N} (b_\mu^\dagger)^{m_\mu} \ket{0},
\end{equation}
where $\ket{0}$ represents the vacuum state. From eq.~\eqref{eq:fermion-lindblad-operator}, the explicit evolution of the matrix elements of the density operator in the Fock basis reads
\begin{widetext}
\begin{align}
    \frac{d}{dt}\rho_{\vec{m}, \vec{n}} = & -{i}\left[\rho_{\vec{m},\vec{n}}\sum_{\mu} \Omega_{\mu}(m_\mu-n_\mu)+\sum_{\mu \neq \nu}V_{\mu \nu}\left\{ m_\mu (1-m_\nu) \rho_{\vec{m}-\vec{e}_{\mu}+\vec{e}_{\nu},\vec{n}}-(1-n_\mu) n_{\nu} \rho_{\vec{m},\vec{n}+\vec{e}_{\mu}-\vec{e}_{\nu}}\right\} \right]\nonumber \\ 
    &-\frac{1}{2}\rho_{\vec{m}, \vec{n}} \sum_{\lambda}\gamma_{\lambda}(m_{\lambda}-n_{\lambda})^2 \nonumber \\
    &+\sum_{\nu} \theta_{\nu} \left[ (1-m_\nu)(1-n_{\nu}) \rho_{\vec{m}+\vec{e}_{\nu},\vec{n}+\vec{e}_{\nu}} - \frac{1}{2}(m_{\nu}+n_{\nu}) \rho_{\vec{m} , \vec{n}} \right] \nonumber \\
    &+\sum_{\nu} \eta_{\nu} \left[ m_{\nu} n_{\nu} \rho_{\vec{m}-\vec{e}_{\nu},\vec{n}-\vec{e}_{\nu}} - \frac{1}{2} (2-m_{\nu}-n_{\nu}) \rho_{\vec{m},\vec{n}} \right]. 
    \label{master equation for fermion}
\end{align}
\end{widetext}
Following the same argument as in Section~\ref{sec:boson} in the ultradecoherence limit, one then obtains
\begin{widetext}
\begin{align}
\rho_{\vec{m},\vec{n}} \approx &  \frac{- i \sum_{ \mu \neq \nu}  V_{\mu \nu}\left\{ {(1-m_\nu)m_\mu}\rho_{\vec{m}-\vec{e}_\mu +\vec{e}_{\nu},\vec{n}}-{(1-n_\mu)n_\nu}\rho_{\vec{m},\vec{n}+\vec{e}_{\mu}-\vec{e}_{\nu}}\right\}}{\frac{1}{2}\sum_\lambda \gamma_\lambda(m_\lambda-n_\lambda)^2+ i\sum_\mu \Omega_{\mu} (m_\mu-n_\mu)} \quad \red{(\vec{m} \neq \vec{n})}.
    \label{eq:fermion-off-diagonal}
\end{align}
\end{widetext}
Observe that eq.~\eqref{eq:fermion-off-diagonal} for fermions can be obtained from eq.~\eqref{eq:bosonic-off-diagonal} for bosons by replacing $(1+m_\nu)$ and $(1+n_\mu)$ with $(1-m_\nu)$ and $(1-n_\mu)$, respectively.
To the first order $\mathcal{O}(1/\gamma_{\mu})$, $\rho_{\vec{m},\vec{n}}$ also vanishes, except when $\vec{n}=\vec{m}-\vec{s}_{\mu\nu}$, where we have
\begin{align*}
    \rho_{\vec{m},\vec{m}-\vec{s}_{\mu\nu}} \approx \frac{- i V_{\mu \nu} {(1-m_\nu)m_\mu}}{i\Omega_{\mu\nu}+\gamma_{\mu\nu}} \left(\rho_{\vec{m}-\vec{s}_{\mu\nu} ,\vec{m}-\vec{s}_{\mu\nu}}-\rho_{\vec{m},\vec{m}}\right),
\end{align*}
in similarity with eq.~\eqref{eq:bosonic-off-diagonal_term-simplified} for bosonic systems.

Then following the same arguments after eq.~\eqref{eq:bosonic-off-diagonal}, one eventually arrives at
\begin{align}
    \frac{d P_{\vec{m}}}{dt} = & \sum_{\mu \neq \nu}  W_{\mu \nu} (1-m_\mu) m_\nu  \left(P_{\vec{m}+\vec{s}_{\mu \nu}} - P_{\vec{m}}\right) \nonumber    \\
    &\quad + \sum_\nu \theta_\nu \left[ (1-m_\nu)  P_{\vec{m}+\vec{e}_\nu} - m_\nu  P_{\vec{m}} \right] \nonumber \\
    &\quad + \sum_\nu \eta_\nu \left[ m_\nu P_{\vec{m}-\vec{e}_\nu} - (1-m_\nu)  P_{\vec{m}} \right],
    \label{eq:classical-master-fermion}
\end{align}
with the same definition of single particle transition rate $W_{\mu\nu}$ as in eq.~\eqref{eq:master-classical} and eq.~\eqref{eq:classical-master-boson}. Again, this equation does not assume or imply that the system can be interpreted as being in a definite Fock state. Such an assumption can be, as above, specifically argued for systems in the ultradecoherence limit. The corresponding consequences are then discussed in Section~\ref{sec:transport}.

\section{Non-equilibrium transport at ultradecoherence}
\label{sec:transport}

As an illustration of the derived emergent stochasticity, we are to discuss some transport phenomena in simple physical setups. Our starting point is the master eq.~\eqref{eq:dynamic_with_creation_and_loss} and eq.~\eqref{eq:classical-master-fermion}, which can be written in a compact form as
\begin{align}
    \frac{d P_{\vec{m}}}{dt} = & \sum_{\mu \neq \nu}  W_{\mu \nu} (1+sm_\mu) m_\nu  \left(P_{\vec{m}+\vec{s}_{\mu \nu}} - P_{\vec{m}}\right) \nonumber    \\
    &\quad + \sum_\nu \theta_\nu \left[ (1+ s m_\nu)  P_{\vec{m}+\vec{e}_\nu} - m_\nu  P_{\vec{m}} \right] \nonumber \\
    &\quad + \sum_\nu \eta_\nu \left[ m_\nu P_{\vec{m}-\vec{e}_\nu} - (1+sm_\nu)  P_{\vec{m}} \right].
    \label{eq:dynamic_with_creation_and_loss_combined}
\end{align}
where $s=1$ is for bosons and $s=-1$ for fermions.

\new{We are to derive the evolution of the local particle density, and discuss the condensation of bosons and the blockage of fermions due to the Pauli exclusion principle. Both phenomena were well discussed in the quantum theory of bosons and fermions, and are sometimes considered to be the manifestation of (macroscopic) quantum mechanical phenomena. Here, we show that they persist in the ultradecoherence limit.
We are also to compute the unique distribution of the time of first arrivals for identical particle in the ultradecoherence regime. Of interest for statistical physics, we also briefly discuss curious different dynamical phases in the exponential growth of bosonic systems.}

\subsection{Evolution of the local particle density}

Let us start by deriving the mean occupation numbers (which describe the local density of particles) from the master equation~\eqref{eq:dynamic_with_creation_and_loss_combined}. The manipulation of the indices during this derivation is straightforward but tedious. To illustrate the procedure, we first ignore the pump and loss terms. Without the pump and loss terms, one multiplies $m_\alpha$ to both sides of eq.~\eqref{eq:dynamic_with_creation_and_loss_combined} and takes the summation over $\vec{m}$ to obtain
\begin{widetext}
\begin{equation}
    \frac{d \mean{m_\alpha}}{d t}
    = \sum_{\mu \ne \nu } W_{\mu \nu} \left( \sum_{\vec{m}} m_{\alpha} (1+sm_{\mu}) m_{\nu} P_{\vec{m} + \vec{s}_{\mu \nu}} - \mean{m_{\alpha} (1+sm_{\mu}) m_{\nu}} \right).
\end{equation}
\end{widetext}
Concentrating on the first term, one would like  to change the summing variable $\vec{m}$ to $\vec{m}+\vec{s}_{\mu \nu}$.
Consider the case of bosons, $s=+1$. This leads to the substitution of $m_\mu$ by $m_{\mu}-1$ and $m_\nu$ by $m_{\nu}+1$. On the other hand $m_\alpha$ changes as $m_{\mu}$ if $\alpha = \mu$, as $m_{\nu}$ if $\alpha = \nu$, and remains constant otherwise. So we have 
\begin{align}
    &\sum_{\vec{m}} m_{\alpha} (1+m_{\mu}) m_{\nu} P_{\vec{m}+\vec{s}_{\mu \nu}} = \mean{m_{\alpha} m_{\mu}  (1+m_{\nu})}
     \nonumber \\
    & -\mean{\delta_{\alpha \mu} m_{\mu} (1+m_{\nu})} +\mean{\delta_{\alpha \nu} m_{\mu} (1+m_{\nu})}.  
\end{align}

For the case of fermions, $s=-1$, one should keep in mind that the occupation numbers can only assume two values: $0$ and $1$. In particular, $m_\mu^2=m_\mu$, $m_\mu (1-m_\mu)=0$. In changing the summing variable $\vec{m}$ to $\vec{m}+\vec{s}_{\mu \nu}$, one can substitute $m_\mu$ by $1-m_{\mu}$ and $m_\nu$ by $1-m_{\nu}$. On the other hand $m_\alpha$ changes as $m_{\mu}$ if $\alpha = \mu$, as $m_{\nu}$ if $\alpha = \nu$, and remains constant otherwise. One then obtains
\begin{align}
    &\sum_{\vec{m}} m_{\alpha} (1-m_{\mu}) m_{\nu} P_{\vec{m}+\vec{s}_{\mu \nu}} = \mean{m_{\alpha} m_{\mu}  (1-m_{\nu})}
     \nonumber \\
    & -\mean{\delta_{\alpha \mu} m_{\mu} (1-m_{\nu})} +\mean{\delta_{\alpha \nu} m_{\mu} (1-m_{\nu})}.  
\end{align}

Therefore, for both bosons and fermions, one arrives at
\begin{align}
    \frac{d \mean{m_\alpha}}{d t}
    =& \sum_{\nu \ne \alpha } W_{\alpha \nu} (\mean{m_{\nu}} -  \mean{m_{\alpha}}) + \nonumber \\    & \quad \sum_{\mu \ne \nu } W_{\mu \nu} \mean{m_\alpha (m_\mu -m_\nu)} .
    \label{eq:mean-intermediate}
\end{align}
The last term in eq.~\eqref{eq:mean-intermediate} also vanishes, as the single particle transition rate is symmetric, $W_{\mu \nu} = W_{\nu \mu}$. 

The loss term and pump term can be manipulated similarly. The final equation for the evolution of the mean occupation numbers then reads
\begin{equation}
    \frac{d \bar{m}_{\mu}}{dt} = \sum_{ \nu \ne \mu} W_{\mu \nu} (\bar{m}_{\nu} - \bar{m}_\mu) + (1+s\bar{m}_{\mu}) \eta_{\mu}- \bar{m}_{\mu} \theta_{\mu},
    \label{eq:mean-evolution-both}
\end{equation}
where we denoted $\bar{m}_{\alpha}=\mean{m_\alpha}$ to simplify the notation.

Equation~\eqref{eq:mean-evolution-both} is somewhat surprising: 
if one ignores the loss and pump, the evolution of the mean occupation numbers follows exactly that of the single-particle master equation~\eqref{eq:master-classical}. 
This is so for both bosons and fermions: the amplification factors of $(1+sm_{\mu})$ are absent from the evolution of mean occupation numbers, \new{as opposed to equation}~\eqref{eq:dynamic_with_creation_and_loss_combined}. 
Intuitively, this can be understood as follows.
The scattering rate from site $\nu$ to $\mu$ is indeed proportional to $(1+sm_\mu) m_{\nu}$. However, the reverse scattering is also proportional to $m_\mu (1+sm_{\nu})$. 
The crucial point is that the microscopic rates of these two processes are symmetric $W_{\mu \nu} = W_{\nu \mu}$.
The actual rate of change of the mean occupation numbers is therefore only proportional to the difference between the forward and reverse transitions, which is $(1+sm_\mu) m_{\nu}-m_\mu (1+sm_{\nu}) = m_{\nu}-m_{\mu}$.

From the above discussion, one can also understand the origin of the amplification factor $1+s \bar{m}_{\mu}$ in the pump terms in eq.~\eqref{eq:mean-evolution-both}. The pump can be considered as a special mode, where scattering from the system back to the pump is not allowed. This creates the microscopic irreversibility and the factor $1+s \bar{m}_{\mu}$ remains.

It should however be emphasised that, even in the absence of pump and loss, equilibrium properties of the identical particles beyond mean occupation numbers are still quite different from that of classically distinguishable particles. As an example, consider the system of $M$ bosons scattering between two modes. That the equilibrium distribution is uniform in Fock space implies that the number of particles in one mode is uniformly distributed from $0$ to $M$. In contrast, for classically distinguishable particles, one expects however a binomial distribution, which can also be approximated by a Gaussian distribution around the mean value of $M/2$.

\subsection{Transport in a one-dimensional diffusion chain}
\label{sec:1d-transport}
Consider a chain of $L+1$ sites arranged linearly, $\mu=0,1,2,\ldots,L$. Particles are injected at site $0$ with rate $\eta$, which diffuse over the chain, and are absorbed  at site $L$ with rate $\theta$. We assume $W_{\mu-1,\mu}=\Gamma$ uniformly for $\mu=1,2,\ldots,L$, which can be referred to as the mobility of the particles; see the sketch in Fig.~\ref{figlattice}.

Consider eq.~\eqref{eq:mean-evolution-both} in the stationary regime, one has
\begin{align}
     \Gamma (\bar{m}_1-\bar{m}_{0}) + \eta (1+ s \bar{m}_0) \label{eq:stationary_0} & = 0,\\
     \Gamma (\bar{m}_{\mu+1}+\bar{m}_{\mu-1}- 2 \bar{m}_{\mu}) \label{eq:stationary_k} & = 0, \\
     \Gamma (\bar{m}_{L-1}-\bar{m}_L) - \theta \bar{m}_L \label{eq:stationary_L} & = 0.
\end{align}
From eq.~\eqref{eq:stationary_k}, one sees that $\bar{m}_{\mu}$ admits a linear profile in the chain, namely,
\begin{equation}
    \bar{m}_{\mu} = \bar{m}_0 + (\bar{m}_1-\bar{m}_0) \mu.
\end{equation}
It then follows that $\bar{m}_L-\bar{m}_{L-1} = \bar{m}_1-\bar{m}_0 = (\bar{m}_L-\bar{m}_0)/L$. Substituting these expressions into eq.~\eqref{eq:stationary_0} and eq.~\eqref{eq:stationary_L}, one obtains
\begin{align}
    \Gamma/L (\bar{m}_L-\bar{m}_{0}) + \eta (1+ s \bar{m}_0) &= 0, \\
    \Gamma/L (\bar{m}_0-\bar{m}_{L})  - \theta \bar{m}_L &= 0,
\end{align}
which can be solved for $\bar{m}_0$ and $\bar{m}_L$. From that, the current flowing through the system can be computed as the rate of particles absorbed at the ending site $L$, $J= \theta \bar{m}_L$, giving
\begin{align}
J = \left[\frac{1}{\eta} - \frac{s}{\theta} - \frac{s L}{\Gamma} \right]^{-1}.
\label{eq:current}
\end{align}

Let us consider different cases. For fermions $s=-1$, we see that the current monotonically increases with increasing the pumping rate $\eta$ at the starting site, the mobility of the particle $\Gamma$, or the absorption rate of the particle $\theta$ at the ending site. In particular, even for $\eta \to \infty$, the current is well-defined and equals $\theta \Gamma/(L\theta + \Gamma)$. This is because the Pauli exclusion principle prevents particles from being injected into the system, when the removal mechanism is not efficient enough and too many particles still accumulate in the system, which can be called Pauli blocking. On the other hand, if either the pumping rate $\eta$, the mobility $\Gamma$ or the loss rate $\theta$ are small, then $J$ is proportional to that parameter, and vanishes if it vanishes.

The case of bosons $s=1$ is rather interesting. Here one sees that, contrary to fermions, the current $J$ monotonically increases when \emph{decreasing} the loss rate $\theta$ as well as the mobility $\Gamma$. This somewhat counter-intuitive behaviour can be easily understood by realising that when the loss rate $\theta$ and/or the mobility $\Gamma$ decrease, particles accumulate in the system. In particular, the starting site accumulates particles, which increases the probability for more particles from the source to be injected into the system. In fact, at
\begin{equation}\label{eqbosonlim}
    1/\eta=1/\theta + L/\Gamma,
\end{equation}
the current becomes infinite, signalling that particles keep accumulating in the system, which never becomes stationary. This resembles the non-equilibrium behaviour of Bose-Einstein condensation, which has been observed in polaritonic systems~\cite{deng2010exciton,carusotto_quantum_2013,byrnes2014exciton}. 
\subsection{Time of first arrivals}
\begin{figure}
    \centering
    \includegraphics[width=0.8\linewidth]{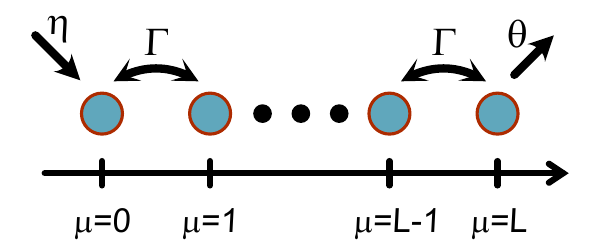}
    \caption{Sketch of the one-dimensional lattice we consider with $L+1$ sites. Particles are injected with rate $\eta$ at the first site $\mu=0$ and absorbed with rate $\theta$ at the last site $\mu = L$. The hopping rate $\Gamma$ is symmetric and uniform along the lattice.}
    \label{figlattice}
\end{figure}

We continue to consider the model discussed in the previous section. The analytical calculation of the time of first arrivals is rather involved and goes beyond the scope of the present work; instead, we turn to Monte-Carlo simulations to get physical insights on its behaviour. We simulate numerically a chain of $L+1=10$ sites, as represented in Fig.~\ref{figlattice}. The parameters of the simulations are expressed with respect to $\Gamma$ which is set to $1$. At each time step, the system will make a transition between two ``neighbouring'' Fock states (meaning, states that differ by only one particle jump to a neighbouring site), and multiple iterations starting from an empty lattice are run to gather statistics. The persistent time before a transition follows an exponential distribution, as we discussed in Section~\ref{sec:time}.

\begin{figure}
    \centering
    \includegraphics[width=\linewidth]{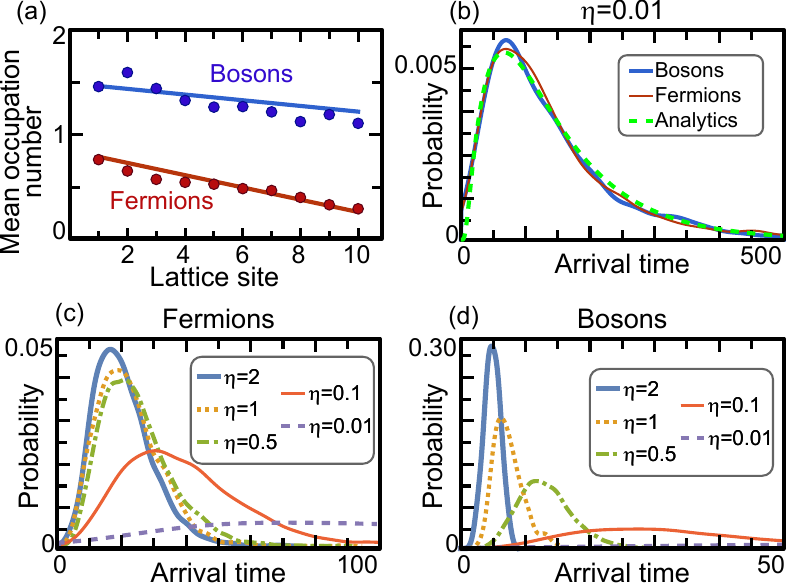}
    \caption{(a) Mean occupation numbers of particles at each lattice site from Monte-Carlo simulations (points) and analytical results (lines). Parameters: $L+1=10$ sites, $\Gamma=1$; $\eta=\theta=0.2$ for fermions (red); $\eta=0.01$ and $\theta =0.02$ for bosons (blue), as an extremely low gain is required for a stationary state to exist, see eq.~\eqref{eqbosonlim}. In simulation, the population is counted in the stationary regime and averaged over $\approx 10^5$ iterations. (b) Results of Monte-Carlo simulations of the probability distribution of arrival times for fermions (red) and bosons (blue) in the low-gain regime $\eta/\Gamma=0.01$. The analytical expression is plotted as a green dashed line. (c,d) Results of Monte-Carlo simulations of the probability distribution of arrival times with different gain values ($\eta/\Gamma=0.01,0.1,0.5,1,2$) for fermions (c), bosons (d). Note that the scales vary between panels.}
    \label{figtime}
\end{figure}

To verify the validity of our calculations, we simulate the chain until reaching a stationary solution and compare the statistics with the analytical calculation of the local particle density from the previous subsection.
The results are plotted in Fig.~\ref{figtime}(a) for bosons ($s=+1$, blue) and fermions ($s=-1$, red). We see an excellent agreement between the Monte-Carlo simulations (points) and the analytical expressions (lines).

Then, we extract from the simulation the first arrival time for different values of gain $\eta$. As we are interested in the first event of particle reaching the site $L$, the parameter $\theta$ plays no role. One is left with two parameters $\Gamma$ and $\eta$, both with dimension of inverse time. Using $\Gamma$ to define the time unit, we consider the distribution of first arrival times dependent on $\eta/\Gamma$.

In Fig.~\ref{figtime}(b), we plot the distribution of the time of arrivals (in units of $1/\Gamma$) for different types of particles in the low-gain regime, $\eta/\Gamma = 0.01$. In this case, the specific particle statistics parameter $s$ should not play a major role because particles rarely interact. This is what we observe also in the simulation results plotted in Fig.~\ref{figtime}(b): for both bosons and fermions, the arrival time distribution follows the same behaviour.
In fact, the distribution of the time of first arrivals for all particle types in the low gain regime coincides and can be analytically computed. Indeed, the first arrival time in the limit of low gain is simply the summation of two independent stochastic times. The former is the time when  a first particle is successfully injected into the system, which is exponentially distributed with exponent $\eta$. The latter is the time that the particle eventually arrives at site $L$, whose tail distribution is given in Appendix~\ref{sec:app-arrivals}.
So, the limiting distribution of the time of first arrivals in the low gain regime can be computed as
\new{
\begin{equation}
    p_T(t) = \frac{2}{2L+1}\sum_{k=0}^{L-1} \frac{\cos^2  (q_k /2)}{(-1)^k \sin (q_k/2)} \frac{ \eta \lambda_k  }{\eta - \lambda_k} \left (e^{-\lambda_k t} - e^{-\eta t} \right ),
    \label{eq:distribution}
\end{equation}}
where $\lambda_k = 2 \Gamma (1 - \cos q_k )$ and $q_k = (2k +1) \pi/ (2L+1)$, as defined in Appendix~\ref{sec:app-arrivals}. The distribution~\eqref{eq:distribution} is plotted as the dashed line in Fig.~\ref{figtime}(b), showing an excellent agreement with the Monte-Carlo simulations.

Next, we increase the ratios $\eta/\Gamma$ to $0.1$, $0.5$, $1$, and $2$. The distribution of arrival times for different particle types are computed for fermions in Fig.~\ref{figtime}(c) and for bosons in Fig.~\ref{figtime}(d). 

The case of fermions, Fig.~\ref{figtime}(c), can be understood intuitively. When the gain is increased, the time of arrival decreases up to a certain limit. Indeed, we see that the distributions for $\eta/\Gamma = 0.5$ and $\eta / \Gamma = 2$ do not differ much, both in mean and variance. This can be understood as follows. When a fermion occupies site $\mu=0$, further particle injection is forbidden. 
Thus even at high gain, the diffusion events of particles at other sites still dominate, which remains inefficient. As a result, the mean arrival time is bounded by a finite nonzero value determined by diffusion processes. The variance of the arrival time is large in all regimes --especially at low gain-- because the contribution from random events dominates.

The arrival time distributions for bosons are plotted in Fig.~\ref{figtime}(d). Increasing the gain leads to a monotonic decrease in arrival time; unlike the fermionic case, the curves for 
$\eta/\Gamma=0.5$ and $\eta/\Gamma=2$ differ clearly in both mean and variance. This reflects the growing dominance of injection events as gain increases. Consequently, particles accumulate, which then causes faster diffusion events. In the infinite-gain limit, the arrival time tends to zero as events become arbitrarily fast.

\subsection{Exponential growth dynamics in infinite dimensional systems}
Interestingly, the exponential growth dynamics of the system of bosonic particles also have different phases. 
In this section, we will not assume a linear chain of modes with nearest neighbour jumps, but rather start with the general dynamics eq.~\eqref{eq:mean-evolution-both}.
To illustrate these non-equilibrium phases, we first assume a uniform pump $\eta_\mu=\eta$, and uniform loss rate, $\theta_{\mu} = \theta$. 
Summing eq.~\eqref{eq:mean-evolution-both} over $\mu$, we have
\begin{equation}
    \frac{d \bar{M}}{dt}= - (\theta-s \eta) \bar{M} + N \eta,
\end{equation}
where $\bar{M}$ is the mean of the total number of particles, $\bar{M}=\sum_{\mu} \bar{m}_{\mu}$, and $N$ is the number of modes. 

For fermions ($s=-1$), $\theta-s \eta > 0$, and the system always reaches equilibrium at particle density of $\eta/(\theta-s \eta)$. On the contrary, for a bosonic system ($s=1$), if $\theta>\eta$, the system becomes stationary with mean density of particles $\bar{M}/N=\eta/(\theta - \eta)$.  
If $\theta < \eta$, i.e., the creation rate is larger than the loss rate of the particles, one has an exponential accumulation of particles into the system, in accordance with the non-equilibrium condensation discussed in Section~\ref{sec:1d-transport}. Interestingly, this exponential growth has different phases in itself. 

For simplicity, we further assume that $W_{\mu \nu} = \Gamma$ (constant) for any $\mu \ne \nu$. Then the evolution of the occupation number at a particular site reads
\begin{equation}
    \frac{d \bar{m}_{\mu}}{dt} = \Gamma \bar{M} - (N\Gamma+\theta - \eta) \bar{m}_{\mu} +\eta.
\end{equation}
One sees that at $\eta= \theta + N \Gamma > \theta$, the growth has a critical behaviour: particles accumulate at all sites with exactly the same speed. In particular, all sites have exponential growth of particles, but the difference of particles at different sites remains constant. This is in contrast with the case $\theta < \eta < \theta + N \Gamma$, where the occupation numbers grow exponentially at any site, but becomes uniform as the difference between different sites actually decays. On the other hand, for $\eta > \theta+ N \Gamma$, the system becomes inhomogeneous, as any difference between occupation numbers at two different sites is exponentially amplified.

\section{Conclusion}

\red{In this work, we have emphasised that the existence of a closed master equation for the probability distribution does not, by itself, justify the use of classical stochastic concepts that rely on conditioning on definite states at intermediate times.}

We argued that such an interpretation becomes meaningful only under additional assumptions, notably a strong separation of timescales leading to ultradecoherence, which allows the system to be regarded as occupying well-defined states over relevant time intervals. Under these conditions, classical stochastic notions such as persistent times and first-arrival-time distributions can be consistently introduced. Worked out examples are given for the stochastic dynamics of a single particle, bosons and fermions. The emergence of classical identical particles is also discussed.

Extensions to more general settings, including situations with incomplete decoherence or memory effects, may help further characterise the limits of classical stochastic reasoning in quantum dynamics. From the foundational viewpoint, our work hints at a possible connection between the time problems in quantum mechanics~\cite{muga_time_2008,muga_time_2009} and the problem of definite outcomes~\cite{zeh_interpretation_1970,zeh_toward_1973,zurek_environment-induced_1982,zurek_decoherence_1991,schlosshauer_decoherence_2007,zurek_decoherence_2003,schlosshauer_decoherence_2005,schlosshauer_elegance_2011,schlosshauer_quantum_2019,kiefer_quantum_2022}, a topic worthy of further investigation.

\begin{acknowledgements}
This work was funded by the Deutsche Forschungsgemeinschaft (DFG, German Research Foundation, project numbers 563437167 and 572827488), the Sino-German Center for Research Promotion (Project M-0294), the German Federal Ministry of Research, Technology and Space (Project QuKuK, Grant No. 16KIS1618K and Project BeRyQC, Grant No. 13N17292), and the Project EIN Quantum NRW.
\end{acknowledgements}

\appendix

\section{The time of first arrivals in classical stochastic processes}
\label{sec:app-arrivals}
One of the examples of the computation of the time distribution of first arrivals can be formulated as follows. Consider a continuous symmetric random walk on a one-dimensional lattice of sites labelled by non-negative integer numbers $0,1,\ldots,L$. The transition rates from sites $k$ to $k\pm 1$ are the same and equal to $\Gamma$. Suppose the particle starts at site $0$, one would like to compute the (stochastic) time it takes for the particle to first arrive at site $L$, which is referred to as the time of first arrivals. 

Crucial to the computation of the first arrival times is to consider the problem for \emph{variable} starting point. Assuming that the system is started at a general site $n$, the arrival time at $L$ is denoted by $\tau_{n}$. We are to compute the cumulative distribution $P(\tau_{n}>t)$.

As the system is initiated at site $n$, after an infinitesimal time $\Delta t$, three possibilities can happen: the system remains at $n$, the system is changed to $n+1$, or to $n-1$, with the corresponding probabilities of $1-2 \Gamma \Delta t$, $\Gamma \Delta t$ and $\Gamma \Delta t$. Thus the probability $P(\tau_{n}>t)$ can be decomposed as
\begin{align}
    P(\tau_{n}>t+\Delta t) = & P(\tau_{n}>t)[1-2 \Gamma \Delta t]  \nonumber \\ & + \Gamma\Delta t P(\tau_{n-1}>t) + \Gamma \Delta t P(\tau_{n+1}>t).
\end{align}
Denoting $P(\tau_{n}>t) = S_n(t)$ and taking the limit $\Delta t \to 0$, one obtains
\begin{equation}
    \frac{d }{dt} S_n (t) =  \Gamma [S_{n-1}(t) + S_{n+1}(t) - 2 S_n (t)].
    \label{eq:arrival-time-equation}
\end{equation}
To model the ending of the chain, one sets $S_{-1}=S_{0}$, so that $\dot{S}_0 (t) = \Gamma [S_{1}(t)-S_{0}(t)]$.
The differential-difference equation then needs to be solved for $S_n (t)$ with the initial condition $S_n (0)=1$ for all $n < L$ and \new{the absorbing boundary condition $S_L(t) = 0$}. 
Techniques to solve such an equation can be found in standard textbooks~\cite{stirzaker_stochastic_2005}. The results read
\begin{equation}
    S_{n}(t) = \frac{2}{2L+1} \sum_{k=0}^{L-1} (-1)^k \frac{\cos (q_k/2) \cos [q_k (n+1/2)]}{\sin (q_k/2)} e^{-\lambda_k t},
\end{equation}
with $q_k = (2k +1) \pi/ (2L+1)$, and $\lambda_k= 2 \Gamma (1 - \cos q_k )$.

It is important here to observe that the derivation of equation~\eqref{eq:arrival-time-equation} itself, as that of the persistent time discussed in Section~\ref{sec:time}, crucially relies on the conditioning of the system on a particular state. 

\section{Example of decoupled dynamics of the diagonal of the density operators}
\label{sec:app-decoupled-diagonals}
We take the example also considered in Ref.~\cite{Joos2003}. This corresponds to the Lindblad master equation~\eqref{eq:lindblad-boson-operator-pump-and-loss} for a single mode with energy $\Omega_{\mu}= \omega$, pumping rate $\eta_{\mu}=\eta$, and loss rate $\theta_\mu =\theta$, 
\begin{align}
    \frac{d \rho}{dt} &= - i \omega [a^\dagger a, \rho] + \theta \left(a \rho a^\dagger - \frac12 \{a^\dagger  a,\rho\} \right) \nonumber \\ 
    & \qquad + \eta \left( a^\dagger \rho a - \frac12 \{a  a^\dagger,\rho\}\right).
    \label{eq:photon-master}
\end{align}
In quantum optics~\cite{WallsMilburn1994QuantumOptics}, this is a standard equation describing a single mode of photon in a cavity with frequency $\omega$ coupled to loss and pump. 
Let $\ket{n}$ denote the Fock state of photon number $n$.
Multiplying the two sides of equation~\eqref{eq:photon-master} from the right with $\ket{n}$ and from the left with $\bra{n}$, one obtains
\begin{equation}
\frac{d P_{n}}{dt} = \theta [(n+1)P_{n+1} - n P_{n}] +  \eta [n P_{n-1}-(n+1)P_n],
\label{eq:master-diagonal}
\end{equation}
where $P_{n} = \bra{n} \rho \ket{n}$ is the probability of having $n$ photons in the system. One sees that the diagonal terms of the density operator in the Fock basis follow exactly a master equation, which is uncoupled from the evolution of the off-diagonals. The evolution of the off-diagonal terms themselves can remain not small and non-trivial~\cite{Joos2003}.
In this case, even though a master equation of the type~\eqref{eq:master-diagonal} is valid, computation based on conditioning the system in a particular photon number such as the time of arrivals is not well-justified; further assumption is generally needed.
\bibliography{rate-equation}

\end{document}